\documentclass[12pt]{article}
\usepackage{authblk}
\usepackage[bookmarksnumbered, colorlinks, plainpages]{hyperref}
\usepackage{amsmath, amsthm, amscd, amsfonts, amssymb, graphicx, color, booktabs}

\usepackage{lineno}

\numberwithin{equation}{section}
\definecolor{email}{rgb}{0.00,0.00,0.84}
\begin{document}
\setcounter{page}{1}

\title{\large \bf 12th Workshop on the CKM Unitarity Triangle\\ Santiago de Compostela, 18-22 September 2023 \\ \vspace{0.3cm}
\LARGE Direct CP violation in $D$ decays at LHCb}

\author[1]{Jolanta Brodzicka (jolanta.brodzicka@cern.ch)}
\affil[1]{Institute of Nuclear Physics PAS, Krak\'ow, Poland}
\maketitle

\begin{abstract}
The discovery of CP violation in charm sector in 2019 was a milestone in searches performed over last decades. The $\Delta A_{CP}$ measured by LHCb is at the upper end of the Standard Model predictions. This triggers many interpretations involving New Physics origins, and calls for more experimental input to clarify the picture. Recent LHCb measurements of CP violation in decays of charmed mesons are presented in these proceedings.
\end{abstract} \maketitle

\section{CP violation in charm decays}

\noindent Breaking of the charge-parity (CP) symmetry occurring in charm decays results in different amplitudes and widths for decays of charmed hadrons and anti-hadrons. This, so-called, direct CP violation (CPV), requires two amplitudes contributing with different weak and strong phases, which within the Standard Model (SM) are realised through tree-level $c \to d, \, s$ transition and penguin $c \to u$ transition. The penguin contribution to charm decays is very small within the SM, due to the CKM matrix elements and the GIM mechanism. Therefore, CPV in charm decays is expected to be suppressed down to $10^{-4} \div 10^{-3}$ level~\cite{DCPV_theory}. It depends mostly on the penguin-amplitude size, which is driven by the level of breaking of the $U$-spin limit based on the equal masses of $s$ and $d$ quarks. In general, direct CPV can differ for different decays, however, it only can occur in singly Cabibbo-suppressed charm decays; the other types receive no penguin contribution. New Physics (NP) contributions can change these SM expectations. 

The most common measurement of the direct CPV is through an asymmetry of decay widths for $D \to f$ and its charge conjugate $\bar{D}\to \bar{f}$, $A_{CP} = \frac{\Gamma(D \to f)-\Gamma(\bar{D}\to \bar{f})}{\Gamma(D \to f)+\Gamma(\bar{D}\to \bar{f})}$. Despite many $A_{CP}$ measurements for numerous decays of primarily charm mesons, the CPV in charm sector was observed only in 2019 via $\Delta A_{CP} = A_{CP}(D^0 \to K^+K^-)- A_{CP}(D^0 \to \pi^+\pi^-)$ using the LHCb data from both Run 1 and Run 2~\cite{dacp_Run2}. The 
measured value, $\Delta A_{CP} = (-15.4 \pm 2.9 )\times 10^{-4}$, is at the upper end of the SM predictions~\cite{dacp_theory}. This triggers many interpretations involving NP origins, and calls for more experimental input to clarify the picture.

\section{$A_{CP}$ in $D^0 \to K^+K^-$ and $D^0 \to \pi^+\pi^-$ decays}

\noindent Measuring $\Delta A_{CP}$ is convenient from an experimental point of view. The difference of the $A_{CP}$ asymmetries is, to a good approximation, equal to the difference of {\it raw} asymmetries between yields of $D^0$ and $\bar{D}^0$ decays, $A_{raw}=\frac{N(D^0)-N(\bar{D}^0)}{N(D^0)+N(\bar{D}^0)}$, while contributions from nuisance asymmetries cancel out. These nuisance asymmetries are, in 
particular, production and detection asymmetries. 
The former is due to different production cross-sections between charmed mesons and anti-mesons in $pp$ collisions, whereas the latter arises from different detection efficiencies between positively and negatively charged hadrons due to their different interactions with the detector material and asymmetries of the detector itself. The detector-related asymmetries are minimised by collecting data with the magnet polarity reversed periodically. Applying fiducial-volume requirements removes soft-pion tracks at the boundaries of the detector acceptance, for which asymmetries are particularly large.  Still, the remaining nuisance asymmetries reach the percent level and, thus, need to be corrected for given that the precision of $A_{CP}$ measurements at LHCb exceeds the $10^{-3}$ level. The LHCb strategy to access $CP$ asymmetries for the  individual channels is to measure $A_{CP}(D^0 \to K^+K^-)$ and disentangle $A_{CP}(D^0 \to \pi^+\pi^-)$ from $\Delta A_{CP}$. 

\begin{figure} [hbt!]
\centering
\includegraphics[width=0.46\textwidth]{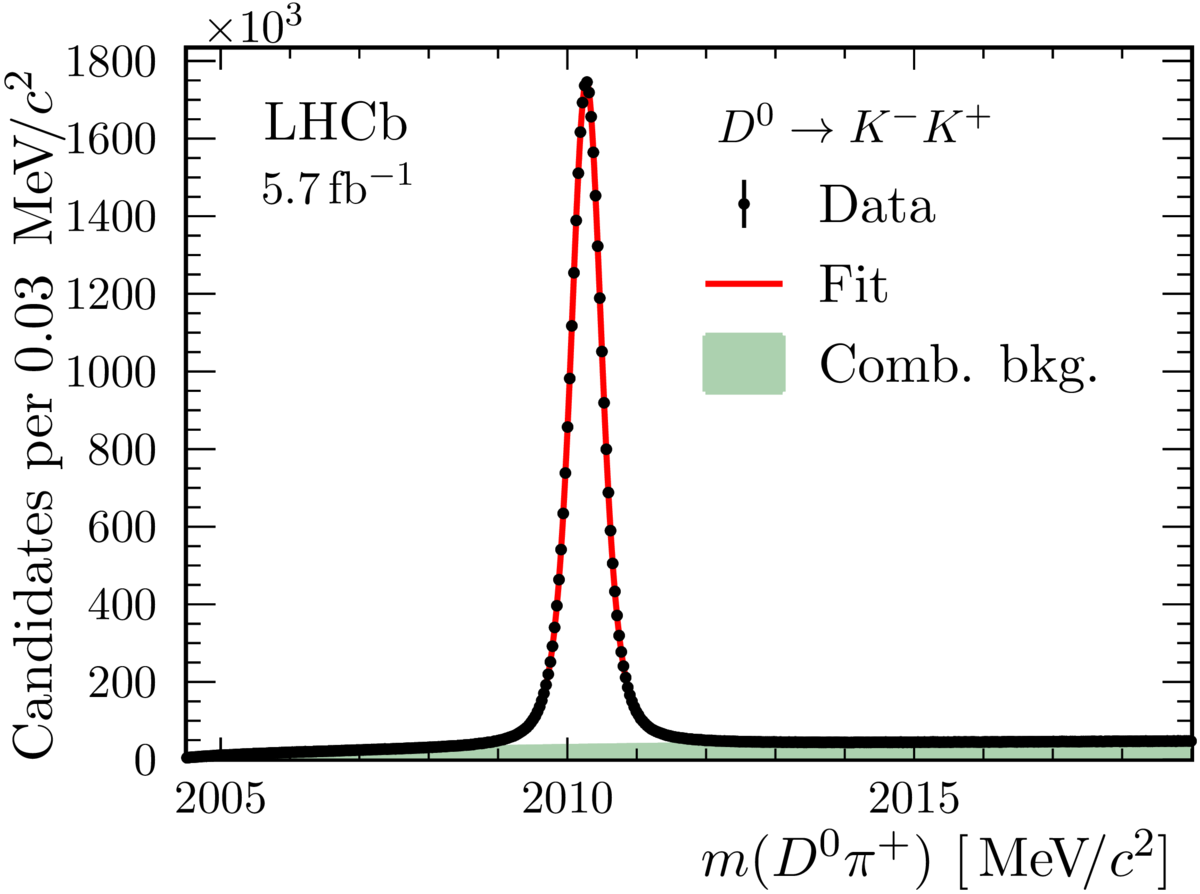}
\includegraphics[width=0.52\textwidth]{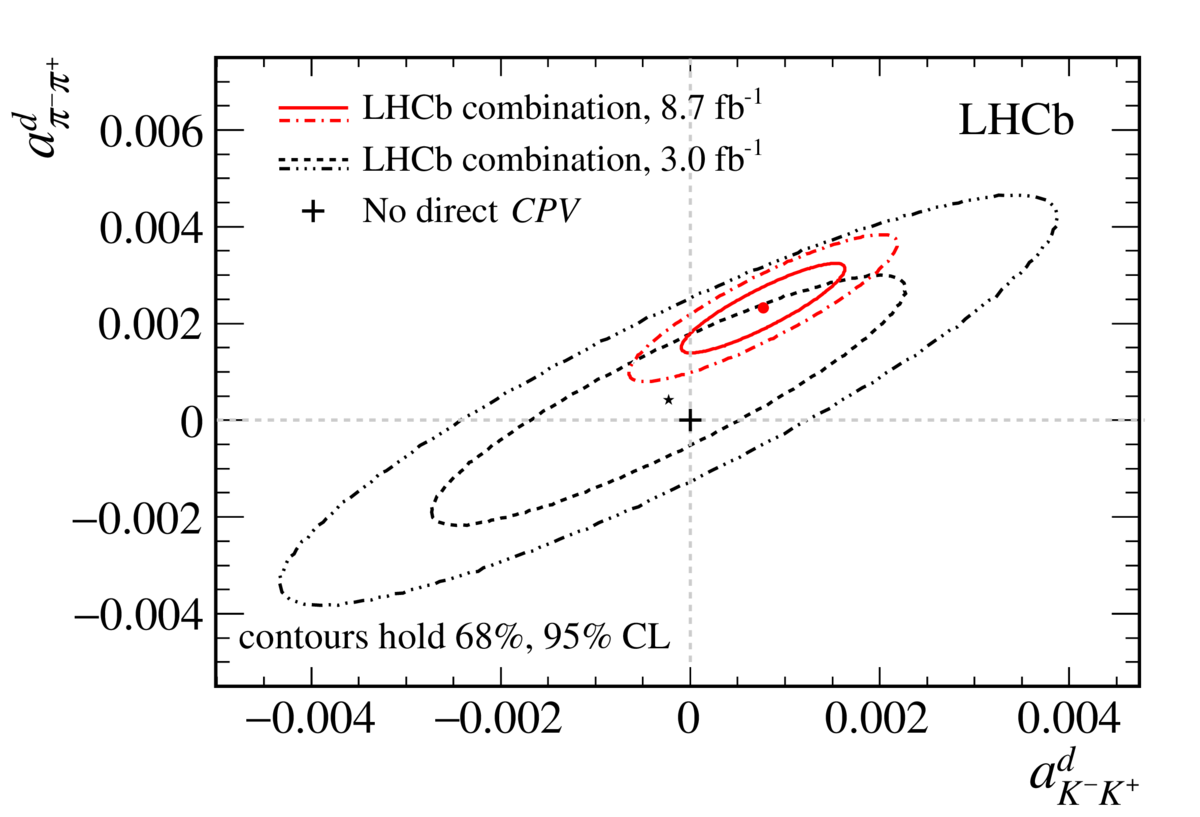}
 \caption{(Left) Distribution of the $D^0 \pi^+$ invariant mass for $D^0 \to K^+K^-$ decays in Run 2 data. (Right) Comparison of direct $CP$ asymmetries in $D^0 \to \pi^+\pi^-$ vs.  $D^0 \to K^+K^-$ decays measured using Run 1 and Run 1+2 data.}
\label{fig:D2KK_mDPi}
\end{figure}

$D^0 \to K^+K^-$ decays originating from $D^{*+} \to D^0 \pi^+_{tag}$ are reconstructed using 5.7~fb$^{-1}$ of the Run 2 data. The electric charge of a {\it tagging} pion $\pi_{tag}$ provides the flavour of a neutral $D$ meson at production: $\pi^+_{tag}$ is associated with $D^0$, while $\pi^-_{tag}$ with $\bar{D}^0$. The $D^0 \pi^+_{tag}$ invariant-mass distribution for reconstructed $D^0 \to K^+K^-$ candidates 
is shown in Fig.~\ref{fig:D2KK_mDPi} (left). It contains about $37 \times 10^6$ signal events with a very high purity. Nuisance asymmetries are removed using high-statistics calibration channels, for which CPV can be neglected. As each of the calibration decays introduces some additional nuisance asymmetries, a set of channels is needed 
so that all nuisance asymmetries cancel each other out  
and $A_{CP}(D^0 \to K^+K^-)$ can be extracted. In addition to the method relied on $D^+$ channels used in the Run 1 measurement~\cite{D2KK_ACP_Run1}, the method based on $D_s^+$ decays is employed in order to increase the precision of the measurement. In these two methods, $A_{CP}(D^0 \to K^+K^-)$ is measured as:
\begin{equation*}
\begin{split}
D^+ \ \textrm{method}:  \  A_{CP}&(D^0 \to K^+K^-) =  A_{raw}(D^0 \to K^+K^-) - A_{raw}(D^0 \to K^-\pi^+) \\
  + & A_{raw}(D^+ \to K^-\pi^+\pi^+) - A_{raw}(D^+ \to \bar{K}^0 \pi^+) + A(\bar{K}^0),  \\
D_s^+ \ \textrm{method}:  \  A_{CP}&(D^0 \to K^+K^-) =   A_{raw}(D^0 \to K^+K^-) - A_{raw}(D^0 \to K^-\pi^+) \\ 
 + & A_{raw}(D_s^+ \to \phi \pi^+)- A_{raw}(D_s^+ \to \bar{K}^0 K^+) + A(\bar{K}^0),  
\end{split}
\end{equation*}
where $A(\bar{K}^0)$ is the asymmetry related to the neutral kaons, and calculated taking into account different cross-sections for $K^0$ and $\bar{K^0}$ interactions with the detector material, as well as the CPV in the $K^0$-$\bar{K^0}$ mixing.  
In general, both production and detection asymmetries are expected to depend on the kinematics of the underlying particle. Therefore, a  multidimensional weighting is performed to match kinematics within pairs of particles in the signal and calibration decays for which cancellation of a given nuisance asymmetry is expected. Combining the two calibration methods gives $A_{CP}(D^0\to K^+K^-) = (6.8 \pm 5.4 \pm 1.6) \times 10^{-4}$. After subtracting a residual contribution due to the CPV related to the $D^0-\bar{D^0}$ mixing, and combining with the Run 1 measurement~\cite{D2KK_ACP_Run1}, the direct CP asymmetry for the $D^0\to K^+K^-$ decays is $a^d_{K^+K^-} = (7.7 \pm 5.7) \times 10^{-4}$, with the statistical and systematic uncertainties combined. The direct CP asymmetry for the $D^0\to \pi^+ \pi^-$ decays, disentangled from the $\Delta A_{CP}$~\cite{dacp_Run2}, is $a^d_{\pi^+\pi^-} = (23.2 \pm 6.1) \times 10^{-4}$ and comprised the first evidence, at $3.8\sigma$, for CPV in individual charm decay. Interestingly, $a^d_{\pi^+\pi^-}+ a^d_{K^+K^-} = (30.8 \pm 11.4) \times 10^{-4} $ is non-zero at $2.7\sigma$, which indicates breaking of the $U$-spin symmetry. 
The comparison of the $a^d_{\pi^+\pi^-}$ vs. $a^d_{K^+K^-}$ measured with the Run 1  and  Run 1+2 data, shown in Fig.~\ref{fig:D2KK_mDPi} (right), demonstrates the sensitivity improvement achieved by LHCb. 

\section{Search for CPV in three-body $D$ decays}

Three-body charm decays mostly proceed via intermediate resonances, which assure strong-phase variation across their phase space. This increases sensitivity to CPV and, thus, makes such decays promising for searching for CPV  localised in regions of their Dalitz distributions. 
Such searches can be performed in a model-independent way, i.e. without modelling resonance amplitudes. 
Phase space of a considered decay is compared with no-CPV hypothesis by means of test statistics.   

$D_{(s)}^+ \to K^-K^+K^+$ decays reconstructed in the Run 2 data have been tested by LHCb~\cite{D2KKK_Miranda} using the binned $S_{CP}$ method. It is based on the observable defined in $i$-th region of a Dalitz distribution as:
\begin{equation*}
\begin{split}
S_{CP}^i=\frac{N^i(D_{(s)}^+) - \alpha N^i(D_{(s)}^-)}{\sqrt{ [\alpha (\sigma_i^2 (D_{(s)}^+) + \sigma_i^2 (D_{(s)}^-))]}}, \ \alpha=\frac{\sum_i N^i(D_{(s)}^+)}{\sum_i N^i(D_{(s)}^-)}, 
\end{split}
\end{equation*}
where $N^i(D_{(s)}^+)$ and $N^i(D_{(s)}^-)$ denote yields of $D_{(s)}^+$ and $D_{(s)}^-$ decays in $i$-th bin, whereas $\sigma_i(D_{(s)}^+)$ and $\sigma_i(D_{(s)}^-)$ are their statistical uncertainties. The factor $\alpha$ is applied to correct for any global asymmetry between the total $D_{(s)}^+$ and $D_{(s)}^-$ yields. $S_{CP}^i$ can be interpreted as significance of the difference between $D_{(s)}^+$ and $D_{(s)}^-$ decay yields in $i$-th region. 
\begin{figure} [hbt!]
\centering
\includegraphics[width=0.45\textwidth,height=0.39\textwidth]{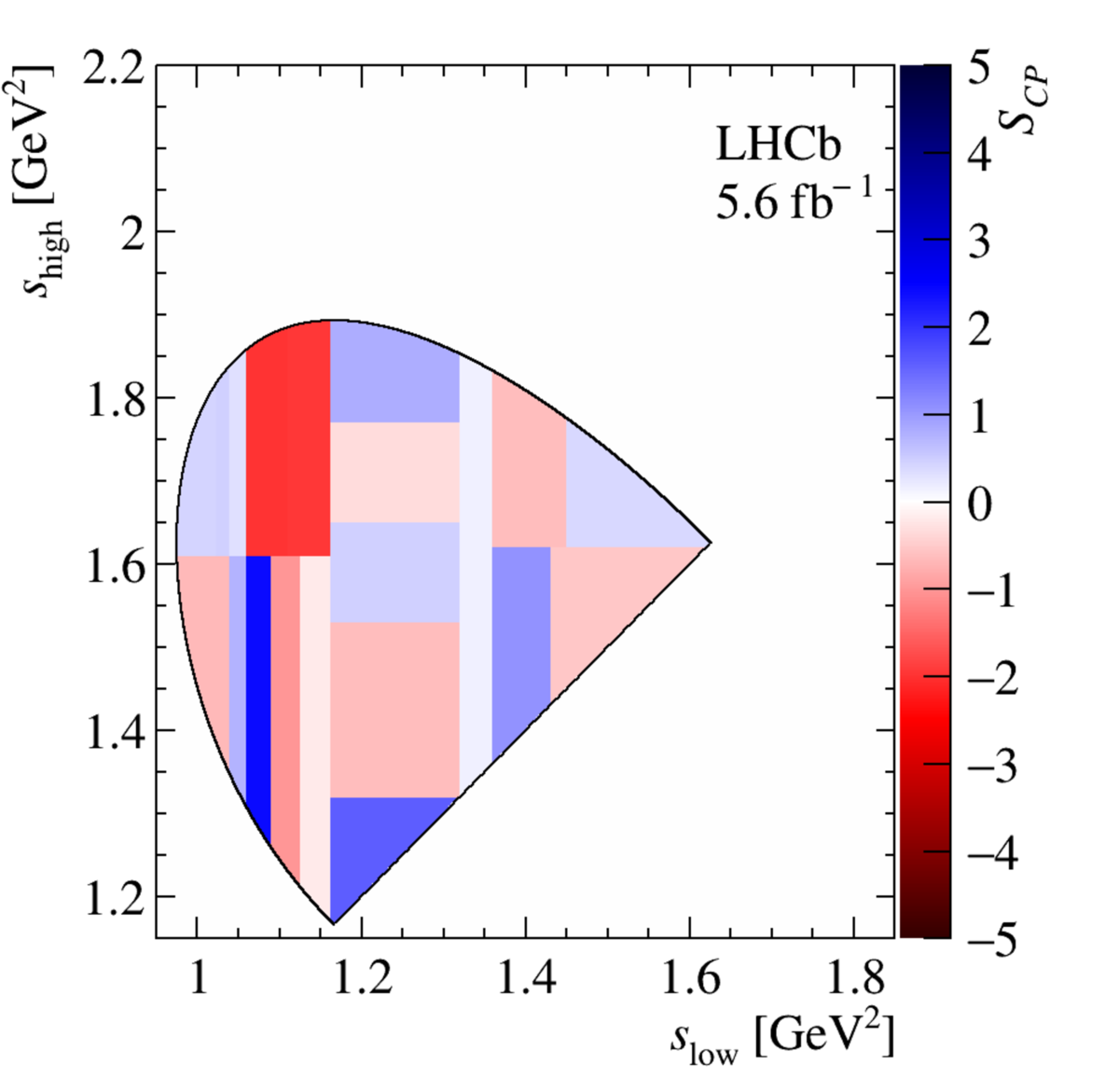}
\includegraphics[width=0.45\textwidth,height=0.39\textwidth]{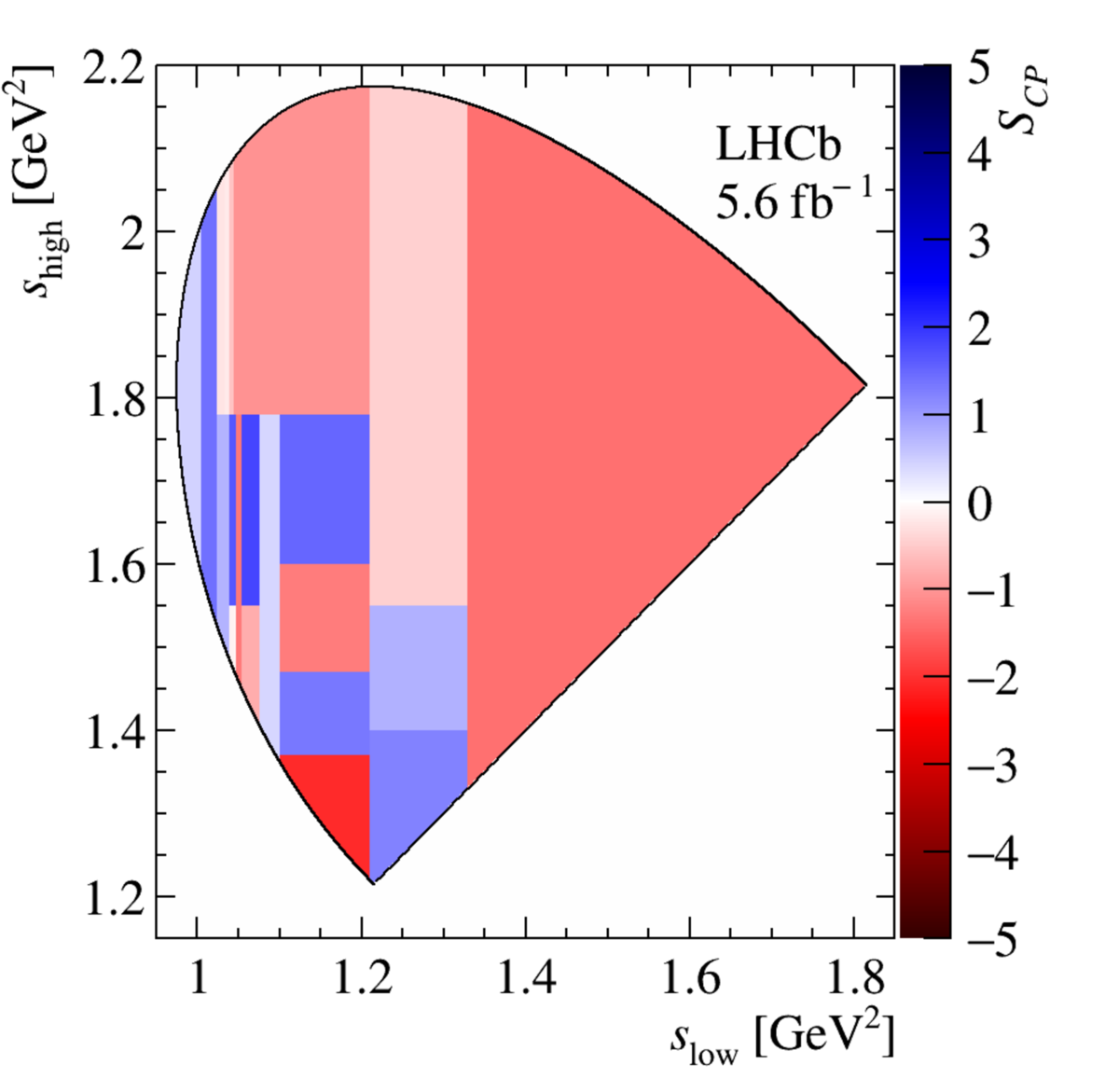}
 \caption{$S_{CP}$ in bins of Dalitz distribution of (left) $D^+ \to K^-K^+K^+$ and (right) $D_s^+ \to K^-K^+K^+$ decays reconstructed in the Run 2 data. The Dalitz plots are folded into invariant-mass squared of the $K^+K^-$ system of higher and lower value, 
 $s_{high}=m_{high}^2(K^+K^-)$ vs. $s_{low}=m_{low}^2(K^+K^-)$.
 }
\label{fig:D2KKK_SCP}
\end{figure}

Figure~\ref{fig:D2KKK_SCP} shows the $S_{CP}^i$ measured in bins of the Dalitz distributions of $D_{(s)}^+ \to K^-K^+K^+$ decays. The optimal binning scheme with 21 bins of similar signal statistics is presented, with the total yields of about $1.0\times 10^6$ $D^{\pm}$ and $1.3\times 10^6$ $D_{s}^{\pm}$ decays. 
For CP symmetric data, the $S_{CP}^i$ distribution is expected to follow a standard normal distribution, and a two-sample $\chi^2$-test is performed with \mbox{$\chi^2 = \sum_i (S_{CP}^i)^2$}, for ndf$=$(number of bins$-1$). The $p$-value for consistency with no-CPV hypothesis is measured to be $32\%$ for $D^+ \to K^-K^+K^+$ and $13\%$ for $D_{s}^+ \to K^-K^+K^+$ decays. Thus, showing no evidence for CPV in these decays.

\begin{figure} [hbt!]
\centering
\includegraphics[width=0.44\textwidth,height=0.32\textwidth]{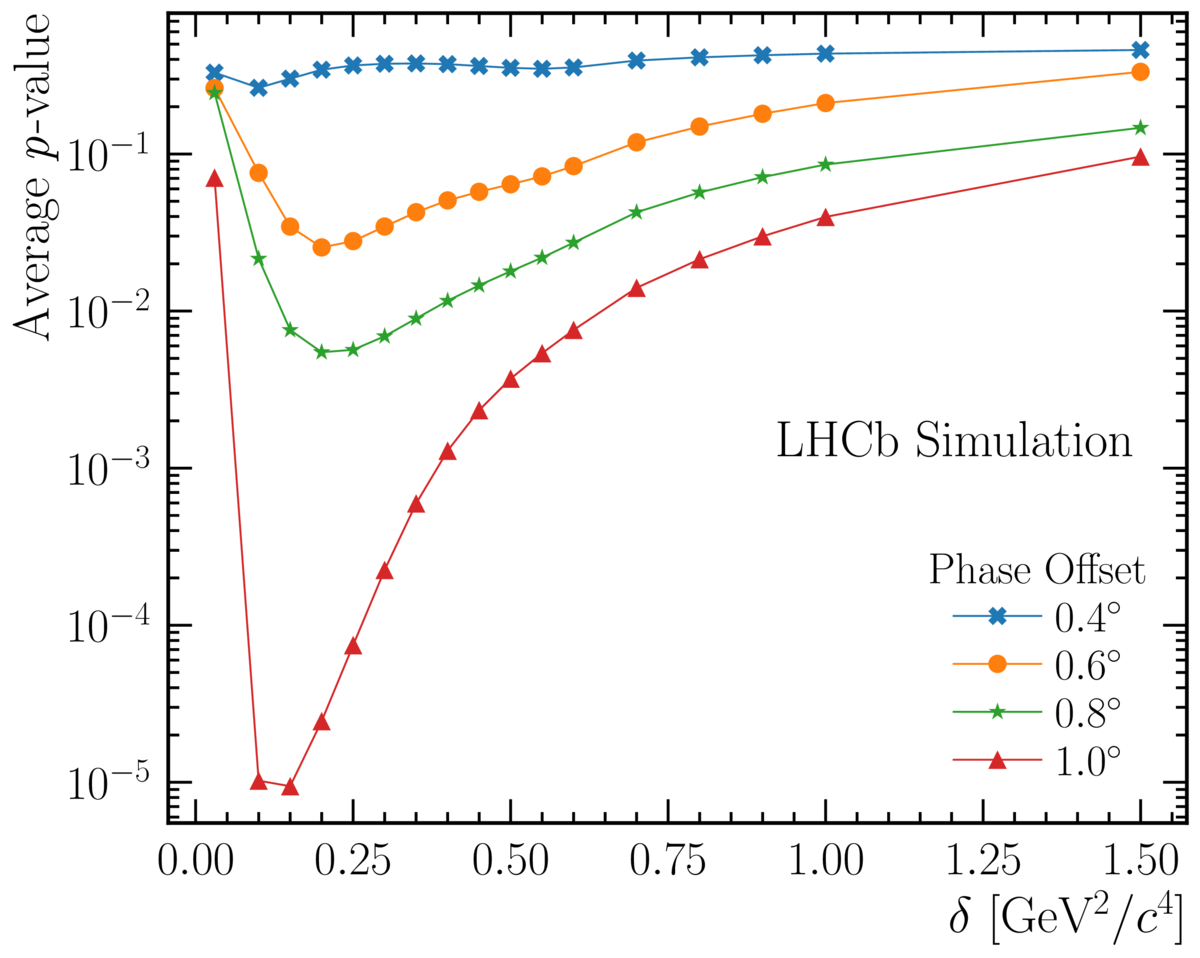}
\hspace{0.4cm}
\includegraphics[width=0.49\textwidth,height=0.34\textwidth]{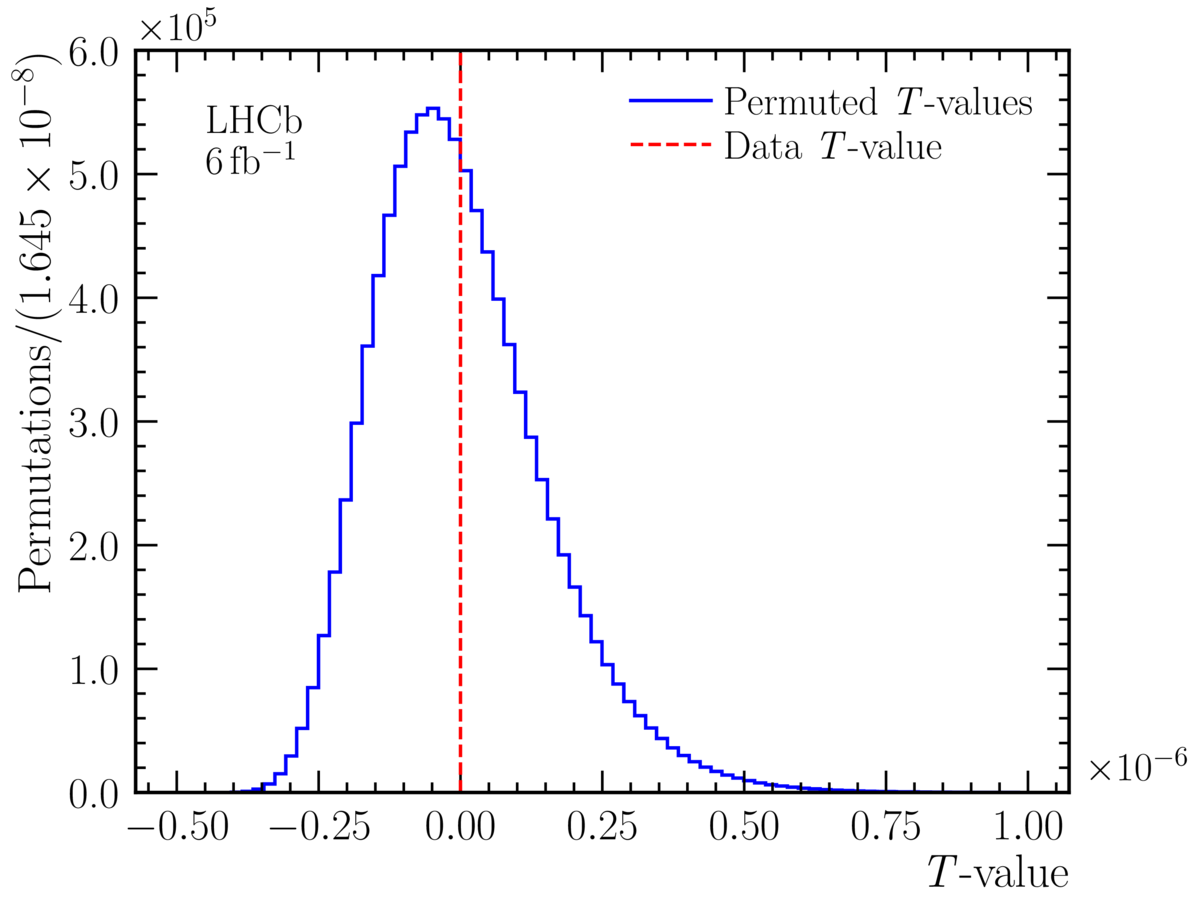}
 \caption{(Left) Optimisation of metric parameter $\delta$ using toy samples of $D^0 \to \pi^+ \pi^- \pi^0$ decays with CPV induced as a phase offset between $D^0 \to \rho^- \pi^+$ and $\bar{D}^0 \to \rho^+ \pi^-$ amplitudes. The optimal value is $\delta = 0.2$~GeV$^2$/c$^4$. (Right) $T$-value distribution measured for $D^0 \to \pi^+ \pi^- \pi^0$ decays reconstructed in Run 2 data.}
\label{fig:D2PiPiPi0_Tvalue}
\end{figure}

To search for CPV in $D^0 \to \pi^+ \pi^- \pi^0$~\cite{D2PiPiPi0_ET_Run2} and $D^{0} \to K_{S}^{0} K^{\pm} \pi^{\mp}$~\cite{D2KsKPi_ET} decays, in the Run 2 data, LHCb has employed the Energy Test. It is an unbinned method for statistical comparison of two distributions, e.g. phase space of $D^0$ and $\bar{D}^0$ decays. Its test statistics $T$ is based on a distance $d_{ij}$ between event pairs $ij$, as:
\begin{equation*}
T= \frac{1}{2n(n-1)}\sum_{i, j \neq i}^{n} \psi_{ij} 
+ \frac{1}{2\bar{n}(\bar{n}-1)}\sum_{i, j \neq i}^{\bar{n}} \psi_{ij}
- \frac{1}{n\bar{n}}\sum_{i, j \neq i}^{n,\bar{n}} \psi_{ij}, 
\end{equation*}
where $n=N(D^0)$ and $\bar{n}=N(\bar{D}^0)$, and the distance function $\psi_{ij}$ is based on the Gaussian metric, $\psi_{ij} =\exp{(-d^2_{ij}/2\delta)}$. Three terms in the above equation correspond to, respectively, an average distance between two events in $D^0$ sample, in $\bar{D}^0$ sample and between events in $D^0$ and $\bar{D}^0$ samples. The distance $d_{ij}$ is calculated in the Dalitz distribution of a considered decay; for $D^0 \to \pi^+ \pi^- \pi^0$ it is \mbox{$d^2_{ij}\!=\![m_i^2(\pi^+ \pi^-) \!\!-\! m_j^2(\pi^+ \pi^-)]^2 \!+\! [m_i^2(\pi^+ \pi^0) \!\!-\! m_j^2(\pi^+ \pi^0)]^2 \!+\!  [m_i^2(\pi^- \pi^0) \!\!-\! m_j^2(\pi^- \pi^0)]^2$}. The metric parameter $\delta$ is optimised using toy samples with CPV induced by introducing either phase or magnitude difference into one of the resonance amplitudes in $D^0$ and $\bar{D}^0$ decays. To make these samples realistic, effects of the phase-space acceptance and background are also included. Figure~\ref{fig:D2PiPiPi0_Tvalue} (left) illustrates such an optimisation performed for the $D^0 \to \pi^+ \pi^- \pi^0$ decays. 

In order to measure $p$-value for no-CPV hypothesis, $T=T_0$ value measured in the data is compared against a $T$ distribution for CP-symmetrized data, in which the $D^0$ or $\bar{D}^0$ flavour is randomly assigned (Fig.~\ref{fig:D2PiPiPi0_Tvalue} (right)). Then $p$-value is calculated as the fraction of events for which $T>T_0$. The $p$-value measured for the $D^0 \to \pi^+ \pi^- \pi^0$ decays is $62 \%$, thus the data are consistent with no-CPV. 
Therefore, despite the signal yield of $2.5 \times 10^6$ events, being by a factor of four larger compared to the Run 1, the small $p$-value of about $2\%$, obtained with the Run 1 data~\cite{D2PiPiPi0_ET_Run1}, has not been confirmed.

$D^{0} \to K_{S}^{0} K^{-} \pi^{+}$ and $D^{0} \to K_{S}^{0} K^{+} \pi^{-}$ decays have different resonance structures~\cite{D2KsKPi_AA}, but both are particularly promising for CPV searches due to presence of amplitudes involving two neutral-kaon states, e.g. $D^0 \to \bar{K}^{*0} K^{0}$ or $D^0 \to \bar{K}^{0} K^{*0}$~\cite{D2K0K0_theory}. Reconstructed signal yields are about $0.9 \times 10^6$ and $0.6 \times 10^6$ events for $D^{0} \to K_{S}^{0} K^{-} \pi^{+}$ and $D^{0} \to K_{S}^{0} K^{+} \pi^{-}$ decays, while the $p$-value for no-CPV hypothesis measured with the Energy Test is $70 \%$ and $66 \%$, respectively. Thus neither of these decays exhibits evidence for CPV. 

\section{Summary}

After the 2019 discovery of CPV in charm sector, LHCb continues its comprehensive studies, in particular of decays of charm mesons. The only evidence for CPV in individual channel is for $D^0 \to \pi^+ \pi^-$ decays, with the direct CP asymmetry of $(23.2 \pm 6.1) \times 10^{-4}$.  Recent searches for CPV in phase space of multibody decays, $D_{(s)}^+ \to K^-K^+K^+$ and $D^0 \to \pi^+ \pi^- \pi^0$, $D^{0} \to K_{S}^{0} K^{\pm} \pi^{\mp}$, performed using model-independent methods, give no indication of CP symmetry breaking. 


{\bf Acknowledgements.} I would like to express my gratitude to the National Science Centre NCN in Poland, for financial
support under the contract no. 2017/26/E/ST2/00934.

\bibliographystyle{amsplain}

\end{document}